\newcommand{\CLASSINPUTtoptextmargin}{2.1cm}
\newcommand{\CLASSINPUTbottomtextmargin}{4.1cm}

\documentclass[conference, a4paper]{IEEEtran}
\usepackage[T1]{fontenc}
\usepackage[latin9]{inputenc}
\usepackage{color}
\usepackage{amsmath}
\usepackage{amsthm}
\usepackage{amssymb}
\usepackage{esint}
\usepackage{verbatim}
\usepackage{graphicx}
\usepackage{stfloats}
\usepackage{subfigure}
\usepackage{dsfont}
\usepackage{bm}
\usepackage{balance}
\usepackage{multicol}
\usepackage{multirow}
\usepackage{tabularray}
\usepackage{makecell}
\usepackage{algorithm}
\usepackage{algorithmic}
\usepackage{mathtools}
\usepackage{array}
\usepackage{subcaption}
\usepackage[numbers,sort&compress]{natbib}

\makeatletter

\definecolor{sblue}{RGB}{0,0,0}
\begin{document}
\title{
 \huge Data-Importance-Aware Waterfilling for Adaptive Real-Time Communication in Computer Vision Applications\\
 
}
\author{Chunmei Xu, Yi Ma$^\dag$, Rahim Tafazolli\\
	{\small 5GIC and 6GIC, Institute for Communication Systems, University of Surrey, Guildford, UK, GU2 7XH}\\
	{\small Emails: (chunmei.xu, y.ma, r.tafazolli)@surrey.ac.uk }}

\maketitle
\begin{abstract}
This paper presents a novel framework for importance-aware adaptive data transmission, designed specifically for real-time computer vision (CV) applications where task-specific fidelity is critical. 
An importance-weighted mean square error (IMSE) metric is introduced, assigning data importance based on bit positions within pixels and semantic relevance within visual segments, thus providing a task-oriented measure of reconstruction quality.
To minimize IMSE under the total power constraint, a data-importance-aware waterfilling approach is proposed to optimally allocate transmission power according to data importance and channel conditions. 
Simulation results demonstrate that the proposed approach significantly outperforms margin-adaptive waterfilling and equal power allocation strategies, achieving more than $7$ dB and $10$ dB gains in normalized IMSE at high SNRs ($> 10$ dB), respectively. 
These results highlight the potential of the proposed framework to enhance data efficiency and robustness in real-time CV applications, especially in bandwidth-limited and resource-constrained environments. 
\end{abstract}

\begin{IEEEkeywords}
Data importance, importance-weighted MSE, waterfilling, real-time communication. 
\end{IEEEkeywords}

\section{Introduction}
Life-transformative applications such as immersive extended reality (XR), telemedicine, autonomous systems, digital twins, and the metaverse are driving rapid advancements in wireless communications and computer vision (CV) \cite{giordani2020toward, wang2023road, semeraro2021digital, wang2023survey}. These applications require unprecedented network performance in terms of data rates, latency, and reliability to deliver real-time, interactive experiences that could redefine healthcare, industrial automation, and personal connectivity. 
Achieving such capabilities will push the boundaries of both communication networks and CV technologies.

For future networks (namely $6$G), this means supporting ultra-high data rates (up to $1$ Tbps), sub-millisecond latency (under 1 ms), and ultra reliability (99.99999\%) \cite{giordani2020toward}. 
Real-time immersive applications like XR cannot afford delays from compression processes, as these would introduce latency that could cause motion sickness or pose risks in telesurgery settings. Consequently, uncompressed data transmission becomes essential, demanding an innovative approach to both network and application-layer design.

CV, meanwhile, is central to these transformative applications, enabling features such as real-time object tracking and virtual-physical mapping through advanced algorithms and deep learning \cite{szeliski2022computer,voulodimos2018deep}. 
However, CV and telecommunications have historically developed along distinct lines. CV emphasizes task-specific metrics, such as detection accuracy, while telecommunications prioritizes data rate and latency. This divergence creates inefficiencies in traditional communication systems, which are often optimized for data fidelity rather than the specific needs of CV applications. Task-oriented semantic communications (SemCom) provide a promising solution by transmitting only the essential ``meaning'' relevant to a task, thus optimizing resource use and aligning with the requirements of CV-driven applications \cite{carnap1952outline, gunduz2022beyond}.

Task-oriented SemCom, often grounded in joint source-channel coding (JSCC) and deep learning, is designed to focus on semantic content critical to a CV task (such as facial features in recognition) rather than transmitting all image details. This approach has the potential to improve both communication efficiency and task-specific performance \cite{xie2021deep, weng2021semantic, erdemir2023generative}. However, several challenges remain. While deep JSCC models can outperform traditional separated coding schemes, they may require analog modulation and often lack generalization across diverse tasks. Recent developments using pre-trained foundation models offer enhanced system compatibility and broaden applicability by training on diverse CV scenarios \cite{xu2024semantic, xu2024generative}.

Despite these advancements, existing SemCom approaches and traditional transmission methods often overlook the varying importance of visual information within CV tasks, leading to inefficient use of radio resources. 
For example, in facial recognition, transmitting facial features accurately is far more important than background details. 
The uniform treatment of data can waste resources and limit CV task efficiency. 
To address this, a paradigm shift in communication strategies is essential: one that captures the hierarchical importance of visual information and aligns resource allocation with task-specific needs. 
This shift brings forth two critical research questions: {\it 1)} How to model a novel metric that reflects the interdependence of CV task requirements, data importance, and telecommunication performance? {\it 2)} How can radio resources be allocated efficiently based on this new metric?

This paper aims to address these questions through the following contributions:

{\it 1)} A novel importance-aware data transmission framework is proposed, where data is partitioned into sub-streams with varying levels of importance. 
    Data importance is modeled through bit positions within pixels and semantic relevance within visual segments, capturing the contribution of each data segment to specific CV tasks. 

{\it 2)} A novel metric termed importance-weighted mean square error (IMSE) is introduced based on the developed importance models. 
This metric provides a task-oriented measure of reconstruction quality, capturing both the task-specific significance of visual information and the interdependence between CV and communication performance.
 
{\it 3)} A data-importance-aware waterfilling approach is developed to minimize IMSE under the total power constraint, yielding the optimal power allocation. The optimal solution depends on data importance in addition to channel conditions,  allocating a larger share of power resources to data with higher importance.

Simulation results demonstrate the superior performance of the proposed approach over margin-adaptive (MA) waterfilling and equal power allocation methods, achieving normalized IMSE gains of more than $7\,\mathrm{dB}$ and $10\,\mathrm{dB}$ at high signal-to-noise ratios (SNRs $ > 10$ dB). Additionally, to reach a satisfactory normalized IMSE performance (i.e. $-26\, \mathrm{dB}$; as discussed in Sec. V), the proposed method reduces the required SNR by $5\,\mathrm{dB}$ and $10\,\mathrm{dB}$ compared to the two baselines, respectively. 
These findings highlight the framework's potential to improve data efficiency and robustness in real-time CV applications, particularly in bandwidth-limited and resource-constrained environments.

\section{{Data-Importance-Aware Communication Model}}\label{sec: system model}
\begin{figure}[tbp]
	\vspace{1em}
	\centering
	\includegraphics[width=1\columnwidth]{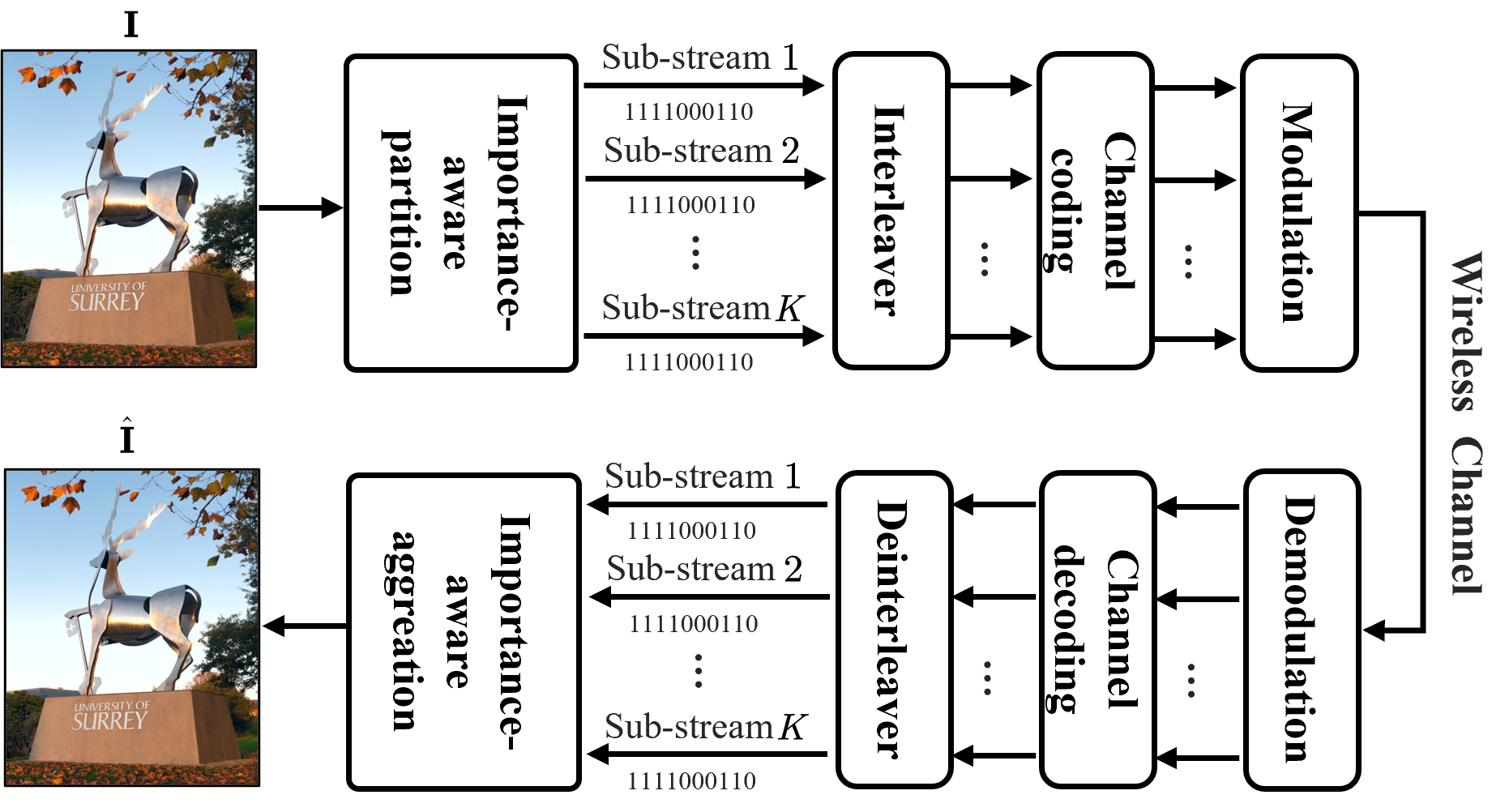}
	
	\caption{The proposed data-importance-aware communication model for CV applications.}
	\label{fig:system model}
\end{figure}
Fig. \ref{fig:system model} illustrates the point-to-point model of data-importance-aware communication in real-time CV  applications. 
The information source is a high-definite image represented as a pixel matrix $\mathbf{I}$ of size $H \times W$.
Each pixel contains multiple color channels, with each channel's pixel values represented by $B$ bits.
For the sake of presentation clarity, this paper focuses on a single color channel, as the principles apply equally to all color channels.

\subsection{Wireless Transmission Model}\label{2a}
Prior to transmission, the pixel matrix $\mathbf{I}$ is partitioned into $K$ bit streams, each with a different level of data importance (see Sec. \ref{sec2b} for details).
Each bit stream is individually passed through a random bit interleaver and then fed into a channel encoder.
After modulation, the information-bearing symbol streams, denoted as $\mathbf{x}_k, _{\forall k\in[1, K]}$, are transmitted through their corresponding sub-channels $h_k$ with transmission power $p_k$. 
The symbol streams received at the receiver, denoted as $\mathbf{y}_k$, are expressed as:
\begin{equation}\label{eq01}
	\mathbf{y}_{k} = h_{k}\sqrt{p_{k}}\mathbf{x}_{k} + \mathbf{v}_{k},~ k=1,...,K,
\end{equation}
where $h_{k}$ is flat block fading, and $\mathbf{v}_k$ is the additive white Gaussian noise with zero mean and varience of $\sigma^2$. 
Given that $\mathbb{E}(\mathbf{x}_k^\mathrm{H}\mathbf{x}_k)=L_k$ (power normalization), the received SNR for the $k$-th stream with length $L_k$ is given by:
\begin{equation}\label{eq02}
	\mathrm{snr}_{k} = \frac{p_{k}\vert h_{k} \vert^2}{\sigma^2},
\end{equation}
where $\mathbb{E}(\cdot)$ stands for the expectation, and $(\cdot)^\mathrm{H}$ for the Hermitian. 
After undergoing a reverse process at the receiver, the pixel matrix is reconstructed as $\hat{\mathbf{I}}$, which is then used for CV-specific tasks.

\subsection{Importance-Aware Data Partitioning}\label{sec2b}
We partition the pixel matrix $\mathbf{I}$ into $K$ sub-streams based on both semantic segment-level importance and sub-pixel-level importance.
\begin{enumerate}
\item[1.] {\bf Semantic segment-level partitioning:} The source image can be semantically divided into $S$ segments using state-of-the-art segmentation models, such as the Segment Anything Model (SAM) \cite{kirillov2023segment}. Each visual segment exhibits varying semantic relevance to the specific CV task; for instance, background segments generally contain less task-critical information than object segments. We model the importance of the $s$-th segment as a non-negative value $\gamma_s \ge 0$, representing its relevance to the CV task, with $\sum_{s=1}^S \gamma_s = 1$.

\item[2.] {\bf Sub-pixel-level partitioning:}  Denote $\mathbf{I}(i,j)$ as the $(i,j)$-th entry of $\mathbf{I}$. It can be represented in polynomial form as:
\begin{equation}\label{eq03}
\mathbf{I}(i,j) = \sum_{b=1}^{B} \mathcal{B}_b(i,j) \cdot 2^{b-1},
\end{equation}
where $\mathcal{B}_b(i,j)\in\{0, 1\}$ represents the $b$-th bit of $\mathbf{I}(i,j)$.
An error in the $b$-th bit (where $\hat{\mathcal{B}}_b \neq \mathcal{B}_b$) introduces an error magnitude of $2^{2(b-1)}$, highlighting that bit position within a pixel significantly affects the error magnitude. Consequently, errors in higher-order bits can severely impact CV task performance, underscoring the need to prioritize accurate transmission for more critical bits. We quantify the importance of the $b$-th bit by its potential error magnitude, modeled as $\gamma_b = 2^{2(b-1)}$.
\end{enumerate}
Together, the semantic segment-level and sub-pixel-level partitioning schemes combine to create $K=(S)(B)$ sub-streams. 

\subsection{Importance-Weighted Mean Square Error}
Conventionally, the error in source reconstruction is measured using the scaled Euclidean norm:
\begin{equation}\label{eq04}
\epsilon=\frac{1}{I}\|\hat{\mathbf{I}}-\mathbf{I}\|^2,
\end{equation}
where $\hat{\mathbf{I}}$ is the reconstructed version of $\mathbf{I}$. $I=(H)(W)$ is the number of pixels of the source image $\mathbf{I}$. 
For a sufficiently large image (e.g., as $I\rightarrow\infty$), the error $\epsilon$ approximates the MSE.  

Following the principle of sub-pixel-level partitioning (specifically as outlined in \eqref{eq03}), the pixel matrix $\mathbf{I}$ can be represented as:
\begin{equation}\label{eq05}
\mathbf{I}=\sum_{b=1}^{B} \mathbf{B}_b \cdot 2^{b-1},
\end{equation}
where $\mathbf{B}_b$ is a binary matrix with $\mathbf{B}_b(i,j)=\mathcal{B}_b(i,j)$ in \eqref{eq03}.
Plugging \eqref{eq05} into \eqref{eq04} results in:
\begin{IEEEeqnarray}{ll}
\epsilon&=\frac{1}{I}\Big\|\sum_{b=1}^{B}2^{(b-1)}(\hat{\mathbf{B}}_b-\mathbf{B}_b)\Big\|^2\label{eq06}\\
&=\frac{1}{I}\Big\|\sum_{b=1}^{B}\sqrt{\gamma_b}(\hat{\mathbf{B}}_b-\mathbf{B}_b)\Big\|^2.\label{eq07}
\end{IEEEeqnarray}
This MSE representation, however, is not well-suited to the optimization task that will be addressed in Sec. \ref{sec3}. 
To address this, we introduce {\bf Assumption 1}: At most one bit out of the B bits in each pixel is incorrectly reconstructed due to communication errors.
Under this assumption, we obtain:
\begin{equation}\label{eq08}
(\hat{\mathbf{B}}_{b1}-\mathbf{B}_{b1})\odot(\hat{\mathbf{B}}_{b2}-\mathbf{B}_{b2})=\mathbf{0}, ~\forall b1\neq b2,
\end{equation}
allowing us to simplify \eqref{eq07} as:
\begin{equation}\label{eq09}
\epsilon=\frac{1}{I}\sum_{b=1}^{B}\gamma_b\|\hat{\mathbf{B}}_b-\mathbf{B}_b\|^2,
\end{equation}
where $\odot$ denotes the matrix Hadamard product. 

Building further on the principle of semantic segment-level partitioning, $\mathbf{B}_b$ is decomposed into $S$ sub-matrices, 
denoted by $\mathbf{B}_b^{(s)}$,
each corresponding to a distinct semantic segment. 
Then, \eqref{eq09} can be further expressed as:

\begin{align}\label{eq10}
\epsilon&=\frac{1}{I}\sum_{b=1}^{B}\gamma_b\sum_{s=1}^{S}\|\hat{\mathbf{B}}_b^{(s)}-\mathbf{B}_b^{(s)}\|^2 \nonumber\\
&=\sum_{b=1}^{B}\gamma_b\sum_{s=1}^{S}\frac{\|\hat{\mathbf{B}}_b^{(s)}-\mathbf{B}_b^{(s)}\|^2 }{I}.
\end{align}
Note that this MSE model does not capture the varying importance of semantic segments, which is crucial for CV-specific tasks.
To address this limitation, we introduce a new task-oriented metric, termed IMSE, as:
{\color{black}\begin{IEEEeqnarray}{ll}\label{eq11}
\mathrm{imse}&=\sum_{b=1}^{B}\gamma_b\sum_{s=1}^{S}\gamma_s \frac{\|\hat{\mathbf{B}}_b^{(s)}-\mathbf{B}_b^{(s)}\|^2}{I_{b,s}} ~\mathrm{s.t.}  \sum_{s=1}^S \gamma_s = 1,
\end{IEEEeqnarray}}where $I_{b,s}$ denotes the number of bits within the sub-matrix $\mathbf{B}_b^{(s)}$. The semantic segment-level importance is reflected by $\gamma_s$ as already discussed in Sec. \ref{sec2b}.
Our resource allocation strategy (in Sec. \ref{sec3}) then seeks to minimize the IMSE through optimum multi-stream (or equivalently multi-sub-channel) power allocation.

{\it Remark 1:} {Assumption 1} introduces a minor approximation to the MSE, which is minimal in scenarios with infrequent communication errors and robust error-correcting mechanisms that effectively limit errors to at most a single bit per pixel.

\section{Optimal Power Allocation with Data-Importance-Aware Waterfilling}\label{sec3}
This section presents an optimal power allocation strategy for data-importance-aware multi-stream transmission, with the goal of minimizing the IMSE.
The IMSE expression in \eqref{eq11} is however not ready to use as it lacks an explicit relationship to the signal power. To address this, we will reformulate the IMSE to incorporate power dependencies, enabling a more effective optimization of power allocation in accordance with the data importance of each stream.

\subsection{Problem Formulation}
Let $e_{b,s}$ represent the semantic segment-level error, which is defined as:
\begin{equation}\label{eq12}
e_{b,s}=\frac{1}{I_{b,s}}\|\hat{\mathbf{B}}_b^{(s)}-\mathbf{B}_b^{(s)}\|^2.
\end{equation}
 The IMSE is then represented as:
\begin{equation}\label{eq12}
\mathrm{imse}=\sum_{b=1}^{B}\gamma_b\sum_{s=1}^{S}\gamma_{s}e_{b,s}, ~\mathrm{s.t.} \sum_{s=1}^S \gamma_s = 1.
\end{equation}
Note that $\mathbf{B}_b^{(s)}$ comprises the bits that form the transmitted sub-stream $\mathbf{x}_{b,s}$. $e_{b,s}$ represents the bit-error rate (BER) of this stream.
Given that each sub-stream is independently coded and decoded, we use the following bit-error-probability (BEP) model to represent the BER \cite{goldsmith2005wireless}:
\begin{equation}\label{eq:BER_SNR}
	\mathcal P^\mathrm{e}_{b,s} = \alpha\exp\left(\beta\mathrm{snr}_{b,s}\right).
\end{equation}Here, $\alpha> 0$ and $\beta< 0$ are  parameters determined by the adopted channel coding and modulation schemes, which can be obtained through data fitting.  
With some tidy-up work, the IMSE form can be represented as:
\begin{equation}\label{eq13}
\mathrm{imse}(p_{b,s})= \sum_{b=1}^{B}\sum_{s=1}^{S}\gamma_b\gamma_{s}\alpha\exp\left(\frac{\beta p_{b,s}|h_{b,s}|^2}{\sigma^2}\right). 
\end{equation}
Then, our optimization problem is formulated as:
\begin{subequations}\label{eq:Prob_segment_bit_level}
	\begin{align}
		 \min_{p_{b,s}} \quad & \mathrm{imse}(p_{b,s})\\
		\mathrm{s.t.} \quad & \sum_{s=1}^{S}\sum_{b=1}^{B}\frac{I_{b,s}}{{\color{sblue}R}\log_2 M}p_{b,s} \le P, \label{eq:cons_power}
	\end{align}
\end{subequations}
where $p_{b,s}$ is the power allocated to each modulation symbol of the $(b,s)$-th sub-stream (see \eqref{eq01}), and $P$ is the total power budget. The term $\frac{I_{b,s}}{\color{sblue}R\log_2 M}$ represents the number of modulated symbols {\color{sblue} with coding rate $R$ and modulation order $M$.}

\subsection{Data-Importance-Aware Waterfilling}
Problem (\ref{eq:Prob_segment_bit_level}) is convex with respect  to $p_{b,s}$, which can be optimally solved using the Lagrange multiplier technique \cite{boyd2004convex}. By introducing the  Lagrange multiplier $\lambda$, the Lagrange function is written by: 
\begin{equation}
	\mathcal L(p_{b,s}, \lambda)\triangleq \mathrm{imse}(p_{b,s}) + \lambda\left( \sum_{s=1}^{S}\sum_{b=1}^{B} \frac{I_{b,s}}{{\color{sblue}R}\log_2 M}p_{b,s} -P\right).
\end{equation}According to the Karush-Kuhn-Tucker (KKT) condition, the optimal solution satisfies: 
\begin{equation}
	\frac{\partial \mathcal L(p_{b,s}, \lambda)}{\partial p_{b,s}}= \gamma_s \gamma_{b}\frac{\partial \mathcal P_{b,s}^{\mathrm{e}}}{\partial p_{b,s}}+ \frac{I_{b,s}\lambda}{{\color{sblue}R}\log_2 M}=0,
\end{equation}where \begin{equation}\frac{\partial \mathcal P_{b,s}^{\mathrm{e}}}{\partial p_{b,s}} = \alpha\beta\frac{\vert h_{b,s} \vert^2}{\sigma^2}\exp\left(\beta\frac{p_{b,s}\vert h_{b,s} \vert^2}{\sigma^2}\right).
\end{equation} 

Since the allocated power cannot be negative, the optimal solution $p_{b,s}^*$ is derived as:
\begin{align}\label{eq:allop_segment_bit_level}
		 &p_{b,s}^*= \left( \frac{\sigma^2}{\beta \vert h_{b,s} \vert^2}\ln \frac{-I_{b,s}\lambda^*\sigma^2}{\alpha\beta {\color{sblue}R}\log_2 M \gamma_s\gamma_b \vert h_{b,s} \vert^2} \right)^+\nonumber\\
		& =  \underbrace{\frac{-\sigma^2}{\beta \vert h_{b,s} \vert^2}}_{W_{b,s}}\left(\underbrace{\ln \frac{-\alpha\beta{\color{sblue}R} \log_2 M}{\lambda^*} }_{H^*_\mathrm{level}} -\underbrace{ \ln \frac{I_{b,s}\sigma^2}{\gamma_s\gamma_b\vert h_{b,s} \vert^2} }_{H_{b,s}} \right)^+,
\end{align}where $(\cdot)^+$ denotes the  $\max(0,\cdot)$ operation.  The optimal Lagrange multiplier $\lambda^*$ is  the solution to the dual problem of (\ref{eq:Prob_segment_bit_level}). 

\begin{figure}[tbp]
	\vspace{1em}
	\centering
	\includegraphics[width=0.95\columnwidth]{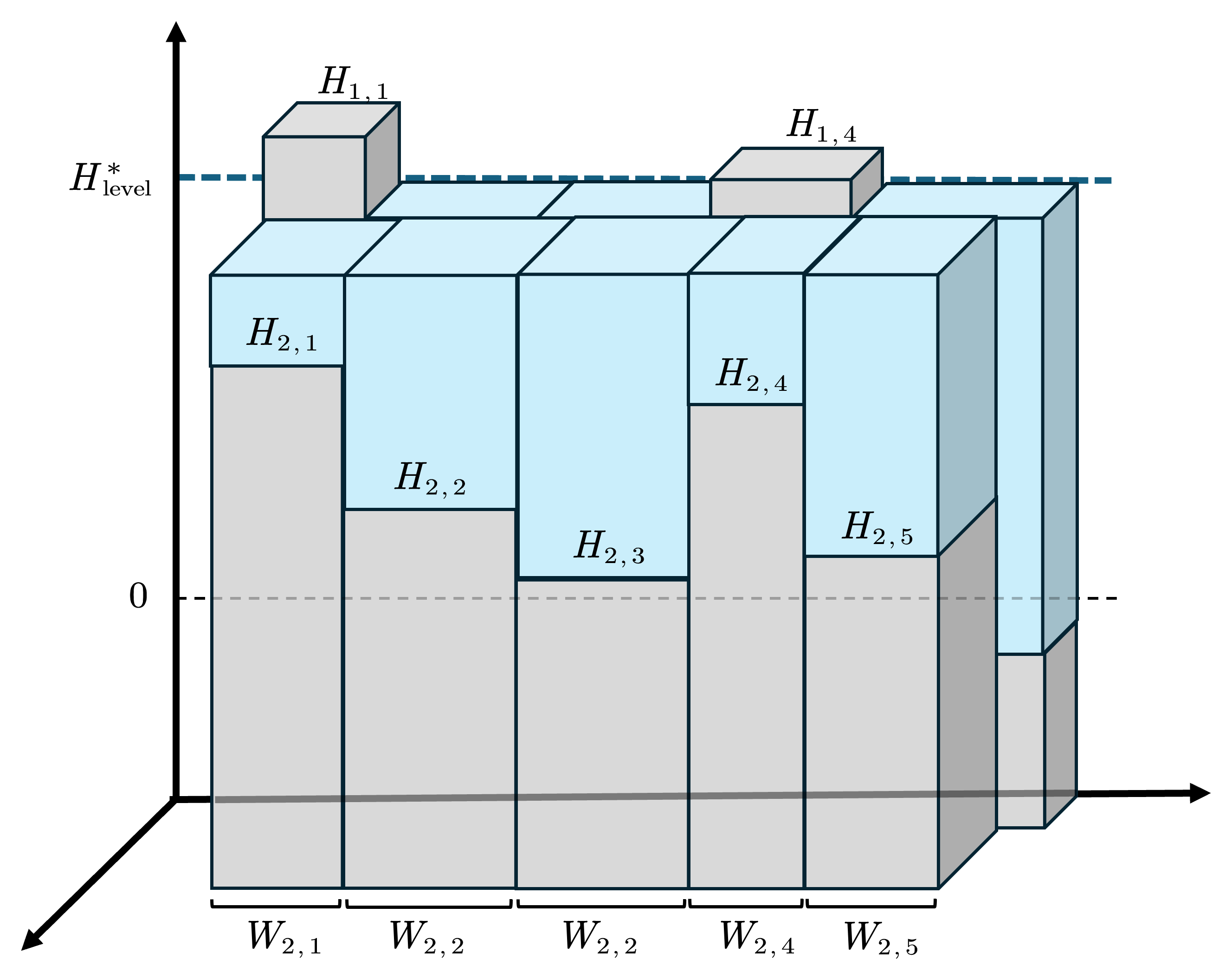}
	
	\caption{An illustration of data-importance-aware waterfilling solution, where $B=5$, $S=2$.}
	\label{fig:Waterfilling-C-BPSS-I}
\end{figure}

This forms the data-importance-aware waterfilling solution as illustrated in Fig. \ref{fig:Waterfilling-C-BPSS-I}. ${W_{b,s}}$ represents the base width, which is jointly determined by the channel condition $\frac{\vert h_{b,s}\vert^2}{\sigma^2}$  of the $(b,s)$-th sub-stream, and the BER parameter $\beta$ corresponding to the channel coding and modulation schemes. $H_{b,s}$ represents the base height, which is determined by the channel condition $\frac{\vert h_{b,s}\vert^2}{\sigma^2}$, importance wights $\gamma_s$ and $\gamma_b$, and the bit length   $I_{b,s}$ of the $(b,s)$-th sub-stream. $H_\mathrm{level}^*$ represents the optimal water level, which is determined by BER parameters $\alpha$ and $\beta$, {channel coding rate \color{sblue}R}, modulation order $M$, and the optimal Lagrange multiplier $\lambda^*$. It satisfies the equality of power constraint (\ref{eq:cons_power}) such that:

\begin{equation}\label{eq:Power_solution}
	\sum_{s=1}^{S}\sum_{b=1}^{B} \frac{I_{b,s}}{{\color{sblue}R}\log_2 M}W_{b,s} \left(H^*_\mathrm{level} - H_{b,s}\right)^+ = P.
\end{equation} $H_\mathrm{level}^*$   can be solved using the incremental search technique. The proposed data-importance-aware waterfilling method to  solve the optimal $p_{b,s}^*$ is summarized in Algorithm \ref{alg:alg3}.  

{\it Remark 2:} the optimal power allocation is jointly determined by both  data importance and channel condition of each sub-stream. $p_{b,s}^*$ is monotonically increasing  with its importance weights $\gamma_b$ and $\gamma_s$, demonstrating that sub-streams carrying more important data receive higher power allocation priority. Moreover, $p_{b,s}^*$ initially increases with the channel condition $\frac{|h_{b,s}|^2}{\sigma^2}$ up to a threshold point, beyond which it decreases. This indicates that power is not necessarily prioritized to sub-channels with exceptionally favorable channel conditions above this threshold.

\begin{algorithm}[t]
	\renewcommand{\algorithmicrequire}{\textbf{Input:}}
	\renewcommand{\algorithmicensure}{\textbf{Output:}}
	\caption{ Data-Importance-Aware Waterfilling.} 
	\label{alg:alg3}
	
	\begin{algorithmic}[1]		
		
		\STATE  Initialize water level $H_{\mathrm{level}}$:\\
		\[H_{\mathrm{level}} = \frac{ P+\sum_{s=1}^{S}\sum_{b=1}^{B}\frac{I_{b,s}}{{\color{sblue}R}\log_2 M} W_{b,s} H_{b,s}}{\sum_{s=1}^{S}\sum_{b=1}^{B}\frac{I_{b,s}}{{\color{sblue}R}\log_2 M}W_{b,s}}\]
		\STATE Initialize $p_{b,s}$ based on (\ref{eq:allop_segment_bit_level})
		
		\STATE \textbf{While} $\big\vert\sum_{s=1}^{S}\sum_{b=1}^{B}\frac{I_{b,s}}{{\color{sblue}R}\log_2 M}p_{b,s}-P\big\vert/P\ge \delta$
		
		\STATE \quad Update water level $H_{\mathrm{level}}$:\\
		\[H_{\mathrm{level}} \leftarrow H_{\mathrm{level}} - \frac{\sum_{s=1}^{S}\sum_{b=1}^{B}\frac{I_{b,s}}{{\color{sblue}R}\log_2 M}p_{b,s}-P} {\sum_{s=1}^{S}\sum_{b=1}^{B}\frac{I_{b,s}}{{\color{sblue}R}\log_2 M}W_{b,s}}
		\]\\	
		\STATE \quad Compute $p_{b,s}$ based on (\ref{eq:allop_segment_bit_level})
		
		\STATE  \textbf{End}
		\STATE Output the optimal solution $p_{b,s}^*$.
	\end{algorithmic}  
\end{algorithm}

\section{Simulation Results}              The simulations consider a point-to-point data-importance-aware communication scenario using  random interleavers, convolutional codes and QAM modulations. Identical channel coding and modulation schemes are applied across all sub-streams, where the coding rate and the modulation order are set to $R=1/2$ and $M=16$, respectively. 
For channels, Rayleigh fading with $h_k\sim\mathcal{CN}(0,1)$ is considered unless otherwise stated, and the noise variance is set to $\sigma^2=1$. { Under these settings, the fitting parameters of BER function in (\ref{eq:BER_SNR}) are $\alpha=0.5123$ and $\beta=-0.2862$. }  The source image is a $640\times 512$ RGB image, where each pixel is represented by $B=8$ bits per color channel. Using the SAM, the image is segmented into three semantic regions: ``stag'', ``base'', and ``background''. These segments are assigned the importance weights of 
$\gamma_{\mathrm{stag}}=0.4975$, $\gamma_{\mathrm{base}}=0.4975$, and $\gamma_{\mathrm{background}}=0.0050$ 
respectively, indicating that ``stag'' and ``base'' are significantly more important than the "background" for a specific CV task. Note that these importance weights depend on specific CV tasks at the receiver, though their measure is beyond the scope of this work.  Table \ref{tab:setup} lists the simulation parameters. For performance comparison, two baselines are considered: 
\begin{itemize}
	\item Equal power allocation: The power is equally allocated to all modulated symbols, regardless of data importance and channel conditions. 
	\item {MA waterfilling \cite{ma2008bit}: The allocated power is obtained  to account for different channel conditions. The objective is to minimize  sum  BERs by treating all sub-streams equally important, leading to the allocated power solution that prioritize the short sub-streams.}
\end{itemize}  

\begin{table}[thbp]
	\caption{Parameter Setup}
	\centering
	\begin{tabular}{|c|c|c|c|}
		\hline
		Parameter  & Value & Parameter & Value\\
		\hline
		Channel coding    & Convolutional code & Coding rate & ${\color{sblue}R=}1/2$ \\
		\hline
		Modulation & $16$-QAM  & Noise variance & $\sigma^2=1$\\
		\hline
		Image size  & $640\times 512$  & Bits depth & $B=8$ \\
		\hline	
	\end{tabular}	\label{tab:setup}
\end{table} 

\begin{figure}[t]
	\centering
	\includegraphics[width=0.995\columnwidth]{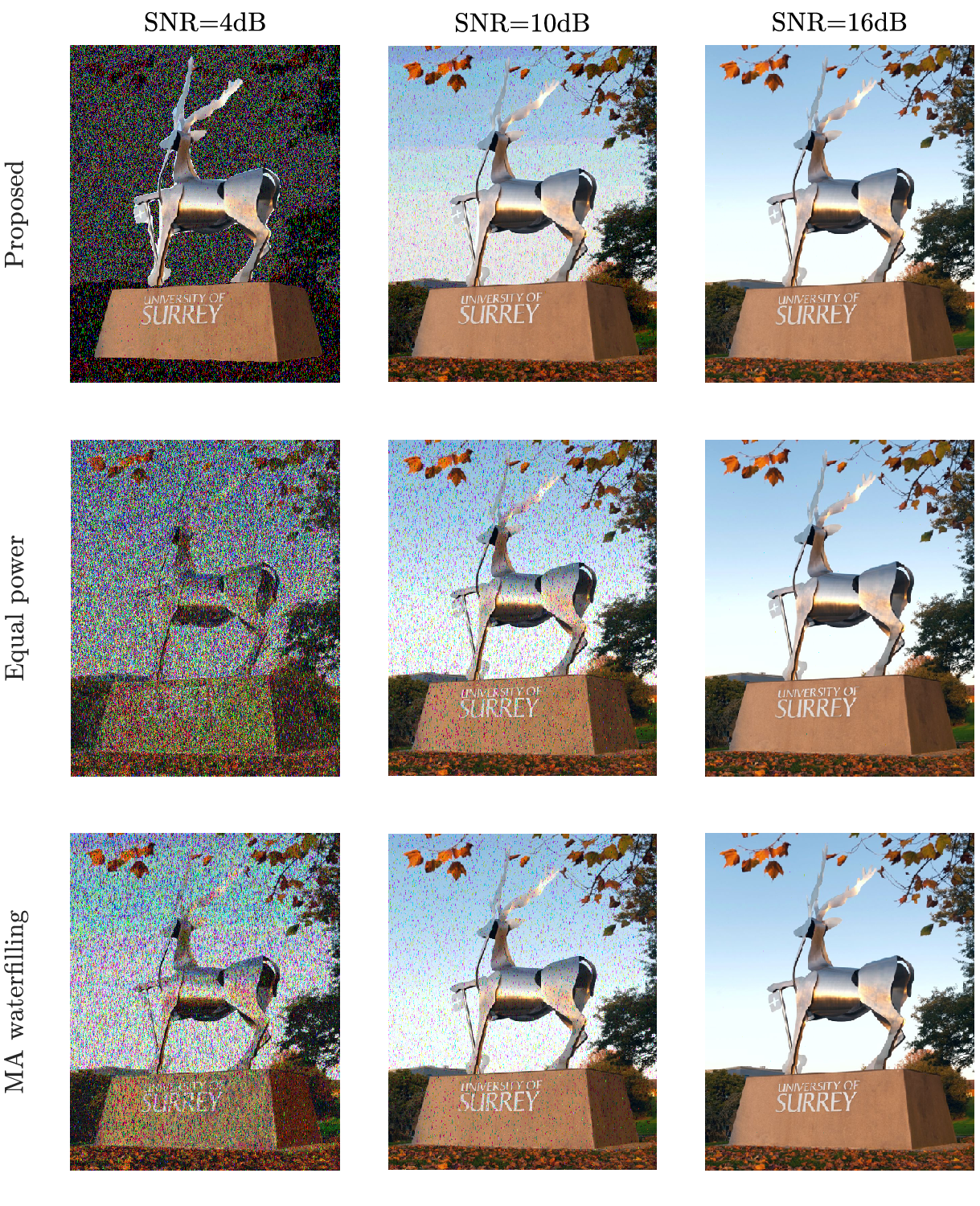}
	\caption{Visual qualities of reconstructed images  with sub-pixel importance weights $\gamma_b=2^{2(b-1)}$, and semantic importance weights $\gamma_{\mathrm{stag}}=0.4975$, $\gamma_{\mathrm{base}}=0.4975$, and $\gamma_{\mathrm{background}}=0.0050$.}
	\label{fig:reconstructed_images_segment_bit_levels}
\end{figure}

\begin{figure}[t]
	\centering	\includegraphics[width=0.995\columnwidth]{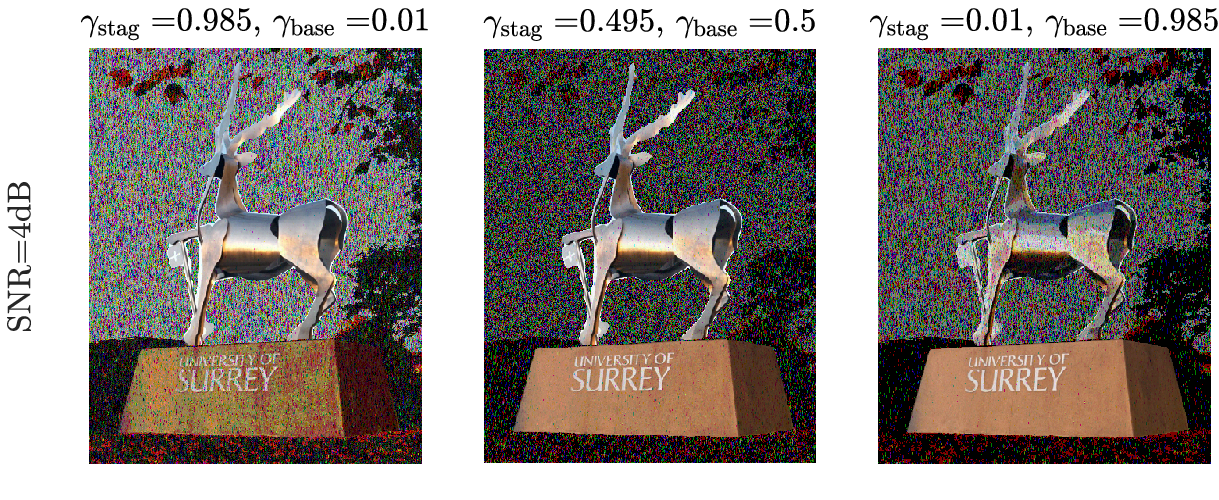}
	\caption{Visual qualities of reconstructed images with bit importance weights $\gamma_b=2^{2(b-1)}$, and varying semantic importance weights of $\gamma_\mathrm{stag}$ and $\gamma_\mathrm{base}$.}
	\label{fig:reconstructed_images_segment_bit_levels_importance}
\end{figure}
To demonstrate the effectiveness of the proposed data-importance-aware communication framework, we examine the visual qualities of the reconstructed images transmitted through AWGN channels without interleaving. Note that under AWGN channels, the power obtained using the MA waterfilling method is affected by the lengths of sub-streams   due to the total power constraint across all sub-streams. 

Fig  \ref{fig:reconstructed_images_segment_bit_levels} compares the visual qualities of reconstructed images using the proposed data-importance-aware waterfilling approach against the two baselines,  where $\mathrm{SNR}$ represents the ratio of the average transmitted signal power to the noise power per symbol. Recall that the ``background'' is assigned by the lowest importance weight ($0.0050$). As SNR increases, the visual qualities of the reconstructed images improve across all methods, with our proposed method demonstrating superior performance compared to the baselines. For example, at $\mathrm{SNR}=4\,$dB, the proposed approach clearly reconstructs key visual elements like ``University of Surrey''   and the  ``stag'', while these remain indistinct using the baseline methods. For the baseline methods, MA waterfilling achieves better visual quality than equal power allocation, as it naturally penalizes longer sub-streams which in this case corresponds to the least important background segment. To examine the impact of semantic weights $\gamma_s$, Fig. \ref{fig:reconstructed_images_segment_bit_levels_importance} shows the reconstructed images at $\mathrm{SNR}=4\,$dB with varying semantic importance weights for ``stag'' and ``base'' segments while fixing the ``background'' weight at $\gamma_\mathrm{background}=0.0050$. The results demonstrate that segments with higher importance weights achieve better visual quality, as they are allocated with a larger sharer of power resource under our proposed data-importance-aware waterfilling method.

\begin{figure}[tb]
	\centering
	\includegraphics[width=0.95\columnwidth]{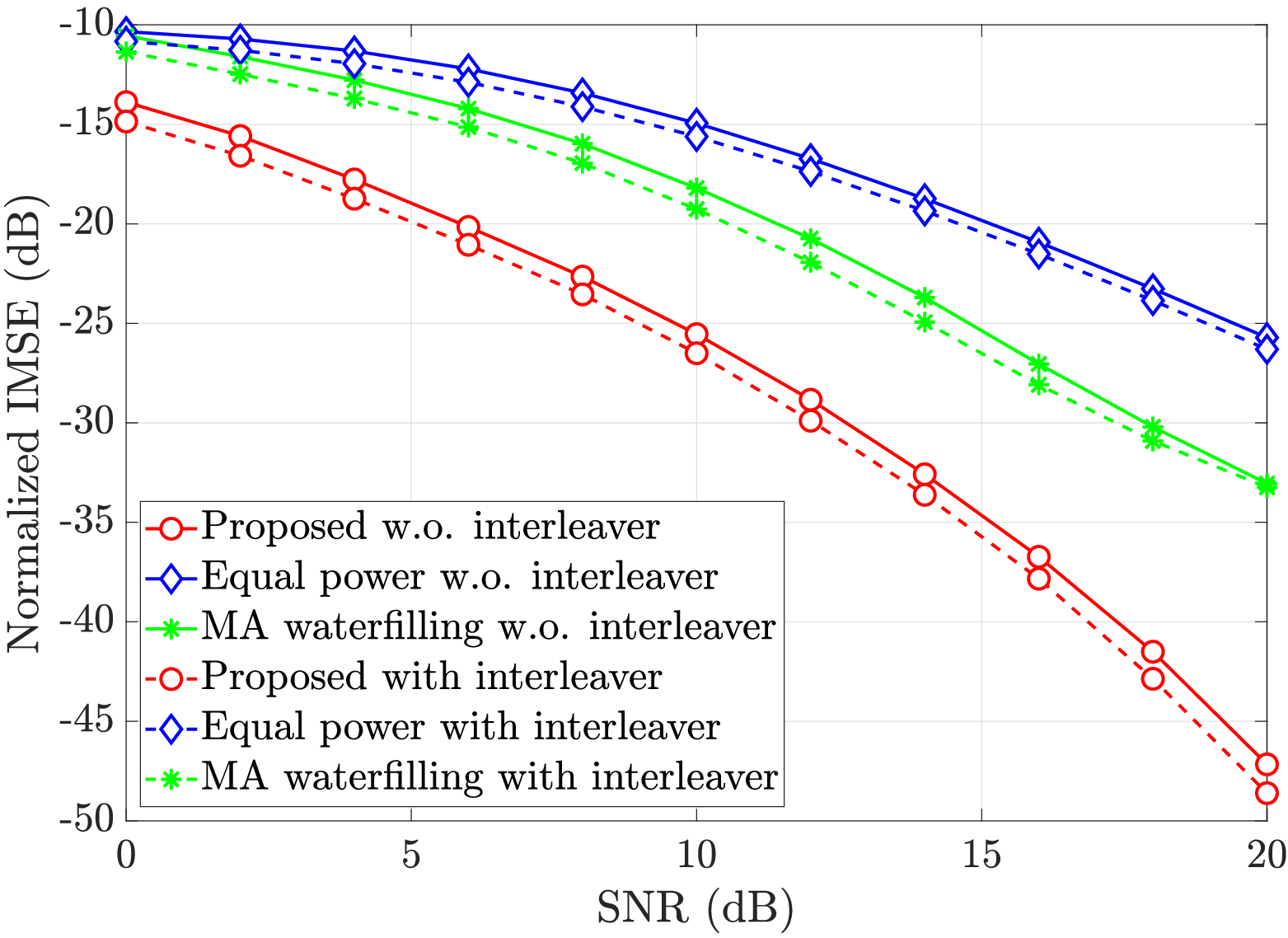}
	\caption{The normalized IMSE performance comparisons with baseline approaches both with and  without interleavers.}
	\label{fig:MSE_VS_SNR_segment_bit}
\end{figure}

The task-oriented reconstruction performance is evaluated by comparing the performance of IMSEs normalized by $\Vert \mathbf{I}\Vert^2/I$. 
Fig. \ref{fig:MSE_VS_SNR_segment_bit} presents the performance comparison of the proposed data-importance-aware waterfilling  approach against  equal power allocation and MA waterfilling methods that treat all sub-streams with equal importance. The results are averaged over 100 channel realizations both with and without implementations of random interleavers. As shown in Fig \ref{fig:MSE_VS_SNR_segment_bit}, while the normalized IMSE decreases with SNR across all approaches, the proposed approach achieve significantly lower normalized IMSE compared to baselines. The MA waterfilling method outperforms the equal power allocation by adapting to channel conditions, rather than distributing power uniformly regardless of channel gains. Moreover, random interleavers further improve the normalized IMSE performance across all approaches by addressing the correlation in sequential pixels, which typically have similar values resulting in consecutive zeros and ones in sub-streams that reduce the error correction capability of convolutional codes. Notably, at high SNRs where $\mathrm{SNR}>10\,$dB, the proposed data-importance-aware waterfilling method achieves more than $7\,$dB and $10\,$dB gains, compared to the MA waterfilling and equal power allocation methods, respectively. Furthermore, to achieve a $-26\, \mathrm{dB}$ IMSE performance, the proposed method reduces the required SNR by $5\, \mathrm{dB}$ and $10\, \mathrm{dB}$ compared to the baselines, respectively. These findings highlight the framework's potential to improve data efficiency and robustness in real-time CV applications, particularly in bandwidth-limited and resource-constrained environments.

\section{conclusion}
This paper presented a novel framework for importance-aware adaptive data transmission for real-time CV applications, where task-fidelity is critical. A novel task-oriented metric and an optimal power allocation strategy were developed  to address the misalignment between data importance and current transmission strategies treating data equally.  This novel metric termed IMSE was introduced to evaluate the task-oriented reconstruction quality by incorporating both semantic segment-level importance and sub-pixel-level
importance importance. To minimize the IMSE under the total power constraint, a data-importance-aware waterfilling approach was developed to optimally allocate power to the transmitted sub-streams, prioritizing data of higher importance while considering channel conditions. Simulation results demonstrated the superior performance of the proposed data-importance-aware waterfilling method over MA waterfilling and  equal power allocation approaches in terms of visual qualities and normalized IMSEs. These results highlight potential of the proposed data-importance-aware transmission framework to improve data efficiency and robustness in real-time CV applications, particularly in bandwidth-limited and resource-constrained environments. 

\section{Acknowledgement}
This work was supported by the U.K. Department for Science, Innovation, and Technology under Project TUDOR (Towards Ubiquitous 3D Open Resilient Network).

\balance

\bibliographystyle{IEEEtran}
\bibliography{reference}

\end{document}